\documentclass[conference]{IEEEtran}

\usepackage{graphicx}
\usepackage{color}
\usepackage{psfrag}

\usepackage{subfig}
\usepackage{booktabs}
\usepackage{amsmath,txfonts}
\usepackage[english]{babel}

\everymath{\displaystyle}
\newcommand{\tabincell}[2]{\begin{tabular}{@{}#1@{}}#2\end{tabular}}

\begin{document}

\title{Energetic Galerkin Projection of Electromagnetic Fields between Different Meshes}

\author{\IEEEauthorblockN{Z.~Wang$^{1}$, Z.~Tang$^{2}$, T.~Henneron$^{2}$, F.~Piriou$^{2}$ and J. C.~Mipo$^{1}$}
\IEEEauthorblockA{$^{1}$Valeo electrical systems, 2 rue Andr\'e Boulle, BP 150, 94017 Cr\'eteil, France
\\$^{2}$L2EP, University of Lille 1, 59655 Villeneuve d'ascq, France  aa
\\zifu.wang@outlook.com}}

\maketitle

\begin{abstract}
In order to project electromagnetic fields between different meshes with respect to the conservation of energetic values, Galerkin projection formulations based on the energetic norm are developed in this communication. 
The proposed formulations are applied to an academic example. 
\end{abstract}


\section{Introduction}

In recent years, the numerical studies of coupled problems are more and more investigated. 
These studies deal with the interactions between different physical phenomena, e.g. electromagnetic - thermal or magnetic - mechanic. 
According to the importance of the interaction, coupled problems can be treated with different strategies.
One possibility is that the study domain is discretized on different meshes for different problems.
In this case, it is necessary to communicate and transfer fields between different meshes. 
In the literature, the concept of Galerkin projection based on the $L^2$ error norm provides a very convenient tool to carry out this transfer \cite{Geuzaine99,Farrell2009,transmag2013}. 
In comparison to the direct interpolation, Galerkin projection enjoys several advantages, especially in terms of precision.
However, using this method, the conservation of the energy is not assured between the original and target meshes \cite{numelec2012}. 
To tackle this issue, formulations deduced from the minimization of the energetic norm can be used. 

In this communication, energetic approaches for Galerkin projection are developed in order to conserve the magnetic energy and electric power between different meshes. 
Firstly, the numerical models developed from the minimization of the energetic norm are presented for magneto-static and eddy current problems. 
Secondly, the obtained formulations are applied to an academic example.

\section{Energetic Galerkin projection formulations}
Given a solution of electromagnetic computation on a source mesh, we aim to project this result onto a target mesh which differs. 
The energetic Galerkin projection formulations are investigated for magneto-static problems as well as eddy current problems. 

\subsection{Magneto-static problems}
In order to solve magneto-static problems, different formulations such as the formulations based on scalar or vectorial potentials can be employed.
As a result, either the magnetic field $\mathbf{H}$ or the magnetic flux density $\mathbf{B}$ is conformed with physical properties.
We denote by $\mathbf{H}_s$ and $\mathbf{B}_s$ the fields obtained on the source mesh and by $\mathbf{H}_t$ and $\mathbf{B}_t$ the fields to be calculated on the target mesh. 
The energetic norms of the interpolation error are defined as:
\begin{equation} 
 \varepsilonup_{\mathbf{H}} = \int_D \mu\left|\mathbf{H}_t-\mathbf{H}_s\right|^2d\tau	
\label{eq_min_h} 
\end{equation}
\begin{equation} 
 \varepsilonup_{\mathbf{B}} = \int_D \frac{1}{\mu} \left|\mathbf{B}_t-\mathbf{B}_s\right|^2d\tau												
\label{eq_min_b} 
\end{equation}
with $\mu$ the linear magnetic permeability.
 
Using Whitney elements in 3D, $\mathbf{H}_t$ and $\mathbf{B}_t$ are discretized in the edge and facet element spaces respectively such that $\mathbf{H}_t=\sum \mathbf w_i^e h_i$ and $\mathbf{B}_t=\sum \mathbf w_i^f b_i$. 
Here $\mathbf w_i^e$ (resp. $\mathbf w_i^f$) and $h_i$ (resp. $b_i$) are the shape functions and the values of $\mathbf{H}_t$ (resp. $\mathbf{B}_t$) associated with the $i-th$ edge (resp. facet). 
To project fields onto a target mesh with respect to the magnetic energy, weak formulations based on the minimization of the energetic norm are developed. 
In (\ref{eq_min_h}), the energetic error norm is minimized when its derivatives with respect to all degrees of freedom are equal to zero, thus for each edge $i$:
\begin{eqnarray}
\frac{\partial}{\partial h_i }\int_{D}\mu\left|\mathbf{H}_{t}-\mathbf{H}_{s}\right|^{2}d\tau =  0 \\
\int_{D}\mu(\mathbf w_i^e \cdot \mathbf{H}_{t}-\mathbf w_i^e \cdot \mathbf{H}_{s})d\tau  =  0 \label{eq_stat_h}
\end{eqnarray}

Finally, the matrix equation to solve can be written:
\begin{equation}
\left[C\right]\left[h\right]=\left[F\right]														\label{eq_cx_f}
\end{equation}
where $C_{ij}=\int_D \mu \mathbf{w}_{i}^e\cdot \mathbf{w}_{j}^e d\tau$, $F_{i}=\int_D \mu\mathbf{w}_{i}^e\cdot \mathbf{H}_s d\tau$ and $[h]$ is the vector of degrees of freedom to be calculated.

A similar demonstration can be applied to (\ref{eq_min_b}) for the projection of $\mathbf{B}_s$ with respect to the magnetic energy. 

\subsection{Eddy current problems}
For eddy current problems, either magnetic or electric harmonic formulations can be used in order to calculate harmonic fields. 
We obtain either $\mathbf{H}_s$ or $\mathbf{E}_s$ conformed with physical properties. 
Thus, depending on the used formulation, the energetic error norm to minimize can be defined \cite{Creuse2012}:
\begin{equation}
\varepsilonup_{\mathbf{H}} = \int_{D}\mu\left|\mathbf{H}_{t}-\mathbf{H}_{s}\right|^2d\tau+\int_{D_{c}}\frac{1}{\sigma\omega}\left|\mathbf{curl}\ \mathbf{H}_{t}-\mathbf{curl}\ \mathbf{H}_{s}\right|^2d\tau
\label{eq_dyna_h}
\end{equation}
\begin{equation}
\varepsilonup_{\mathbf{E}} = \int_{D_{c}}\frac{\sigma}{\omega}\left|\mathbf{E}_{t}-\mathbf{E}_{s}\right|^2d\tau+\int_{D}\frac{1}{\mu}\left|\frac{\mathbf{curl}\ \mathbf{E}_{t}-\mathbf{curl}\ \mathbf{E}_{s}}{\omega}\right|^2d\tau 
\label{eq_dyna_e}
\end{equation}
with $\sigma$ the electrical conductivity and $\omega$ the pulsation. 

In order to project fields onto a target mesh with respect to the magnetic energy in $D$ and electric power in conducting domain $D_c$, weak formulations can be obtained by the minimization of the energetic norm.
Equation (\ref{eq_dyna_h}) gives rise to matrix equation:
\begin{equation}
(\left[C\right]+\left[C_e\right]) \left[h\right]=\left[F\right]+\left[F_e\right]													\label{eq_cx_f_dyna}
\end{equation}
where $[C]$ and $[F]$ are the same matrix as in magneto-static problems, $C_{eij}=\int_{D_c} \frac{1}{\sigma \omega} \mathbf{curl\ w}_{i}^e\cdot \mathbf{curl\ w}_{j}^e d\tau $ and $F_{ei} = \int_{D_c} \frac{1}{\sigma \omega} \mathbf{curl\ w}_{i}^e\cdot \mathbf{curl\ H}_s d \tau$. 
Here $[C]+[C_e]$ is a positive-definite matrix. 

A similar matrix system can be obtained from (\ref{eq_dyna_e}) in order to project $\mathbf{E}_s$.

\section{Application}
\begin{figure}[ht]
\centering
\includegraphics[width=3.6cm]{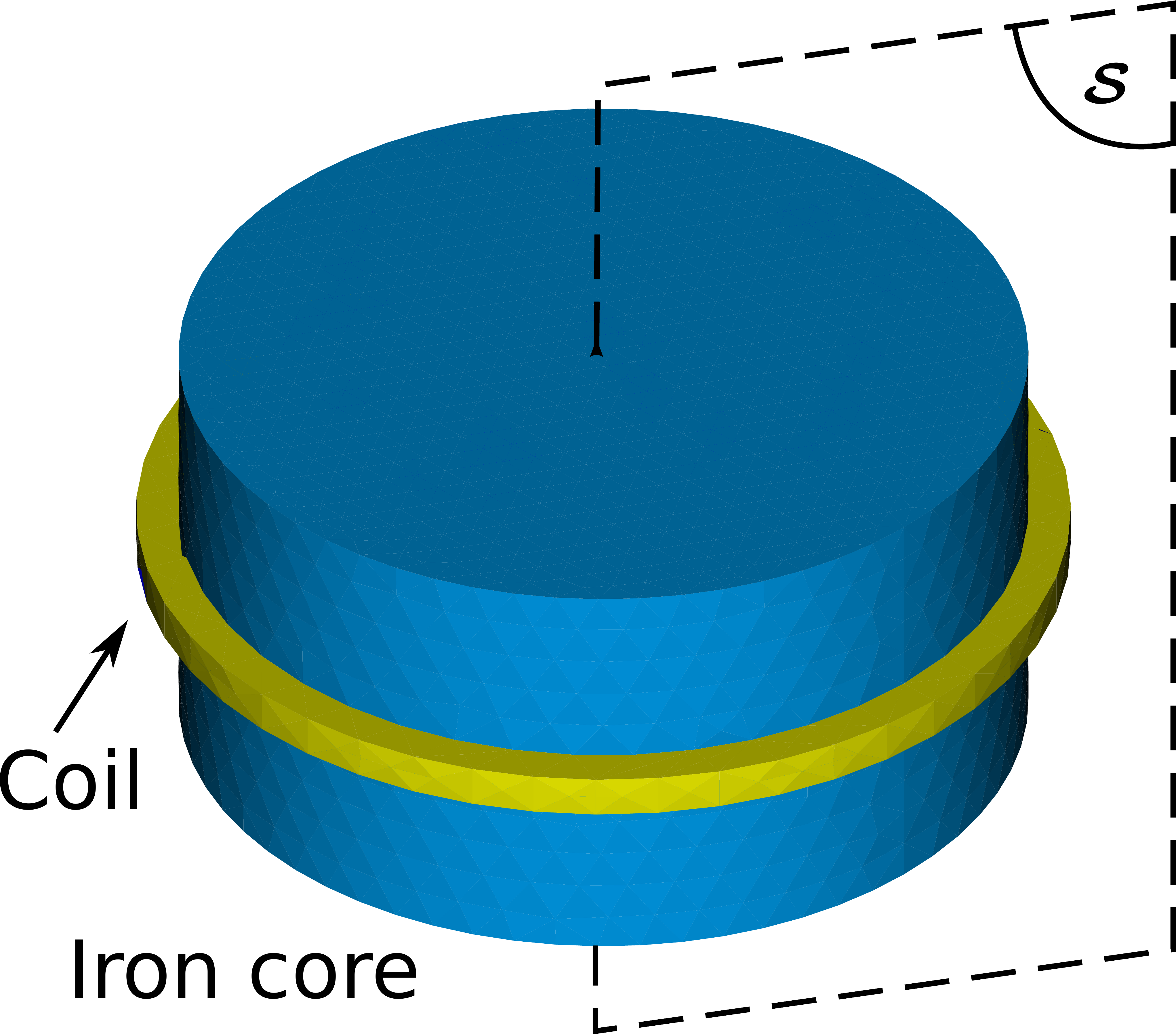}
\caption{Geometry of the example}							\label{fig_geo}
\vspace{-1.5em}
\end{figure}

The proposed projection formulations are applied in an academic example. 
This example is composed of a magnetic core cylinder and an excitation coil (Fig. \ref{fig_geo}). 
Two meshes are considered: Ms (306K elements) and Mt (60K elements). 
In order to carry out projections, Ms is used as the source mesh for the first computation and Mt is considered as the target mesh for the projection. 

\begin{figure}[!t]
\begin{centering}
\subfloat[Energetic Galerkin projection]{\begin{centering}
\includegraphics[bb=0bp 150bp 800bp 900bp,clip,width=4cm]{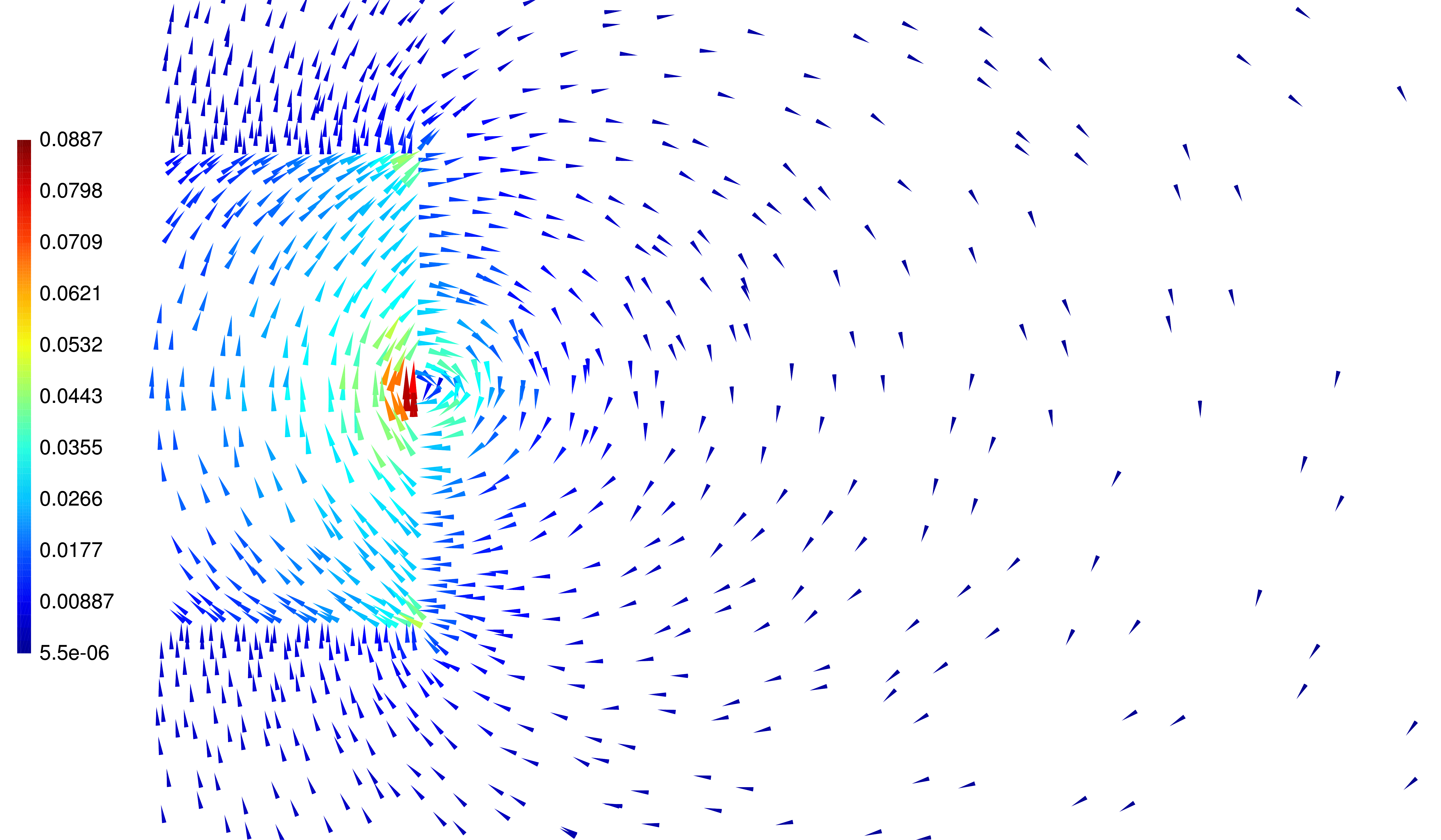}
\par\end{centering}
}
\subfloat[$L^2$ Galerkin projection]{\begin{centering}
\includegraphics[bb=0bp 150bp 800bp 900bp,clip,width=4cm]{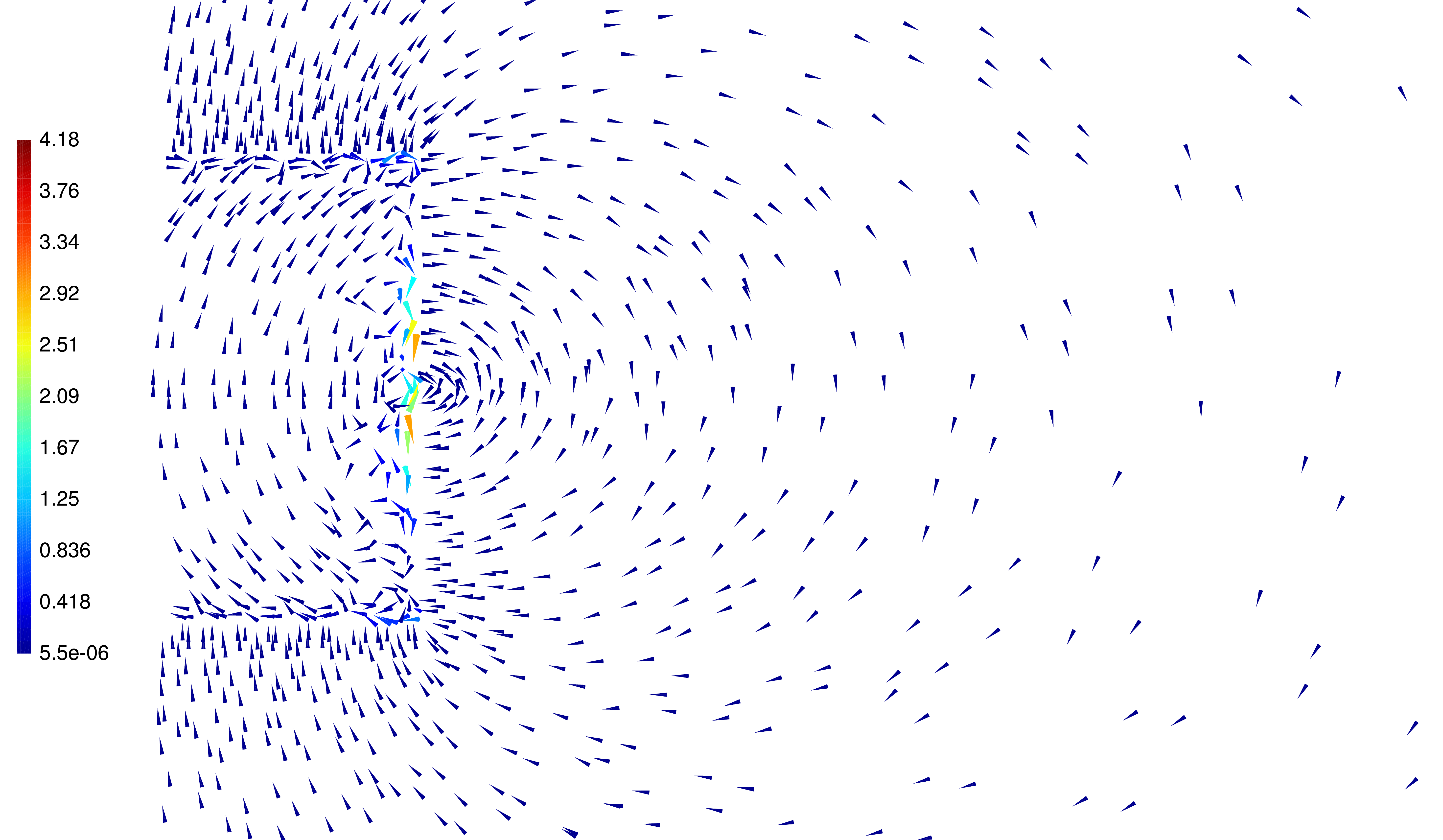}
\par\end{centering}
}
\par\end{centering}
\caption{Distribution of $\mathbf{B}$ (T) on a clipping plane, calculated from the projected $\mathbf{H}_t$ field. }					\label{fig_b_stat}
\end{figure}

Firstly, a constant current is applied in the coil.
The magneto-static problem is solved on Ms or Mt using the scalar potential formulation. 
The obtained field $\mathbf H_s$ on Ms is then projected to Mt using (\ref{eq_cx_f}). 
The corresponding magnetic flux density on the clipping plane (S in Fig. \ref{fig_geo}) is presented in Fig. \ref{fig_b_stat}a.
In order to illustrate the advantage of the projection method, the field $\mathbf H_s$ is also projected using a classical $L^2$ norm projection (Fig. \ref{fig_b_stat}b). 
In comparison to the energetic projection, the $L^2$ projection fails to provide a correct distribution of $\mathbf B$ at the boundary of the cylinder. 
Table \ref{tab_com_stat} presents the max values of fields and the magnetic energy calculated on Ms and Mt alone for reference, and then from Ms to Mt projections.
With the formulation deduced from the minimization of the energetic norm, the magnetic energy and the max values of fields are close to the reference results (calculated on Mt alone).  

Secondly, a sinusoidal current is applied as excitation. 
In this case, eddy currents appear in the conductor cylinder. 
The magneto harmonic problem is solved with the magnetic formulation. 
Table \ref{tab_com_dyna} presents the reference results (calculated on Ms and Mt alone) for the max values of fields, the magnetic energy in the domain and the ohmic losses in the cylinder.
As the $L^2$ projection fails to conserve the magnetic energy and the ohmic losses, only the results obtained using the energetic projection are shown.
They are obtained using (\ref{eq_cx_f_dyna}).
In this example, the magnetic energy as well as the ohmic losses are well conserved using the energetic Galerkin projection.

\begin{table}[!]
\caption{Validation of the energetic projection approach  (magneto-static problem)} 						\label{tab_com_stat}
\begin{centering}
\begin{tabular}{cccc}
\toprule 
 &\tabincell{c}{magnetic energy\\ ($mJ$)} & \tabincell{c}{$|\mathbf{H}|_{max}$\\ ($kA/m$)} & \tabincell{c}{\textbf{$|\mathbf{B}|_{max}$}\\ ($T$)}\tabularnewline
\midrule
calculated on Ms  & 8.01 & 44.1 & 0.116\tabularnewline
calculated on Mt & 8.25 & 49.0 & 0.096\tabularnewline
\midrule 
$L^{2}$ proj. Ms$\rightarrow$Mt& 10.28 & 42.9 & 4.17\tabularnewline
energetic proj. Ms$\rightarrow$Mt& 7.32 & 45.3 & 0.089\tabularnewline
\bottomrule
\end{tabular}
\par\end{centering}
\end{table}

\begin{table}[!]
\caption{Validation of the energetic projection approach (eddy current problem)}						\label{tab_com_dyna}
\begin{centering}
\begin{tabular}{cccccc}
\toprule 
 &\tabincell{c}{magnetic \\energy ($mJ$)} & \tabincell{c}{ohmic losses\\ ($\mu W$)} & \tabincell{c}{$|\mathbf{H}|_{max}$\\ ($kA/m$)} & 
 \tabincell{c}{\textbf{$|\mathbf{J}|_{max}$}\\ ($kA/m^2$)} \tabularnewline
\midrule 
calculated on Ms & 8.05 & 104 & 44.9 & 
 21.1\tabularnewline
calculated on Mt & 8.18 & 104 & 49.7 &
 20.2\tabularnewline
\midrule 
\tabincell{c}{energetic proj. \\ Ms$\rightarrow$Mt}& 7.54 & 98 & 46.8 &
 19.6\tabularnewline
\bottomrule
\end{tabular}
\par\end{centering}
\vspace{-1.5em}
\end{table}

\bibliographystyle{IEEEtran}
\bibliography{biblio}

\end{document}